\documentclass{article}

\usepackage{arxiv}

\usepackage[utf8]{inputenc} 
\usepackage[T1]{fontenc}    
\usepackage{hyperref}       
\usepackage{url}            
\usepackage{booktabs}       
\usepackage{amsfonts}       
\usepackage{nicefrac}       
\usepackage{microtype}      
\usepackage{graphicx}

\title{Digital Twins}

\author{
  Dirk Hartmann\\
  Corporate Technology\\
  Siemens Corporation\\
  Princeton, NJ 08540, USA \\
  \texttt{hartmann.dirk@siemens.com} \\
   \And
  Herman Van der Auweraer \\
  Siemens Industry Software NV\\
  3001 Leuven, Belgium \\
  \texttt{herman.vanderauweraer@siemens.com} \\
}

\begin{document}
\maketitle

\begin{abstract}
Digital Twins are one of the hottest digital trends. In this contribution we will shortly review the concept of Digital Twins and the chances for novel industrial applications. Mathematics are a key enabler and the impact will be highlighted along four specific examples addressing Digital Product Twins democratizing Design, Digital Production Twins enabling robots to mill, Digital Production Twins driving industrialization of additive manufacturing, and Digital Performance Twins boosting operations. We conclude the article with an outlook on the next wave of Digital Twins, Executable Digital Twins, and will review the associated challenges and opportunities for mathematics.
\end{abstract}


\section{Introduction}\label{sec:introduction}

Complexity in today’s industry is exploding. New production methods, miniaturization of electronics, novel sensor technologies, and last but not least the internet of things (IoT) have let to many disruptive development. Ultimately, this leads to more and more complex products. On the one hand, this offers unique opportunities e.g. in terms of efficiency or autonomy of components, products, and complex systems. On the other hand, it challenges today’s design, engineering, operation, and service paradigms mostly focusing on manual expert interaction, which can hardly, if at all, handle this enormous complexity.
 
At the same time digitization changes everything everywhere. With the rise of new technology trends, such as AI Foundations, Intelligent Things, Cloud to Edge, or Immersive Experiences \cite{panetta2017gartner}, many of today’s paradigms can be expected to be disrupted. Along these new technology trends are also Digital Twins, see e.g. \cite{nafems2018hype}. The Digital Twin concept has been originally introduced in 2003 by Michael Grieves \cite{grieves2014digital} and first put to public by the NASA in 2012 \cite{glaessgen2012digital}. Digital Twins integrate all (electronic) information and knowledge generated during the life-time of a product, from the product definition and ideation to the end of its life. They bridge the virtual and real world with the goal to model, understand, predict and optimize their corresponding real assets\footnote{For a more precise definition c.f. Section \ref{sec:digitaltwins}}.

Today, Digital Twins are considered so important to business, that they were named three years in a row among Gartner’s Top 10 Strategic Technology Trends \cite{cearley2017gartner,kerremans2018gartner,kerremans2019gartner}. It is widely accepted that Digital Twins lead to high savings along the entire life-cycle and at the same time will allow for novel services such as on-site diagnostics, prescriptive maintenance or operation optimization \cite{van2018simulation,boschert2016digital}. Ultimately the are expected to be a good for themselves selves enabling complex ecosystems \cite{boschert2018next}.

Within this article we will  review the concept of Digital Twins and relate the concept with opportunities for the mathematical sciences. In particular we will provide a set of examples where advanced mathematical methods have made a difference. Furthermore, we will outline challenges laid down by Digital Twins expecting to spur new mathematical research with impact for industrial applications \cite{krull2019passion}.

Since the concept of Digital Twins is not new, we start with a short history of simulation (Section \ref{sec:history}) before we introduce the concept of Digital Twins (Section \ref{sec:digitaltwins}). We than show it‘s opportunities along the four industrial examples of early design support, manufacturing parallel simulations, a specific manufacturing planning task, as well as operation-parallel simulations (Section \ref{sec:examples}). In all these cases, Digital Twins have allowed for paradigm shifts in their respective domains. Albeit these example prove the success of Digital Twins already today, we believe that they are just taking off. Thus we introduce the next generation concept of Executable Digital Twins in the following (Section \ref{sec:executabledigitaltwins}) and conclude the with the associated challenges.

For a very concise and more detailed review of on Digital Twins we would in particular like to refer to \cite{rasheed2019digital}.

\section{A Short History of Simulation}\label{sec:history}

\begin{figure}
    \centering
    \includegraphics[width=15.0cm]{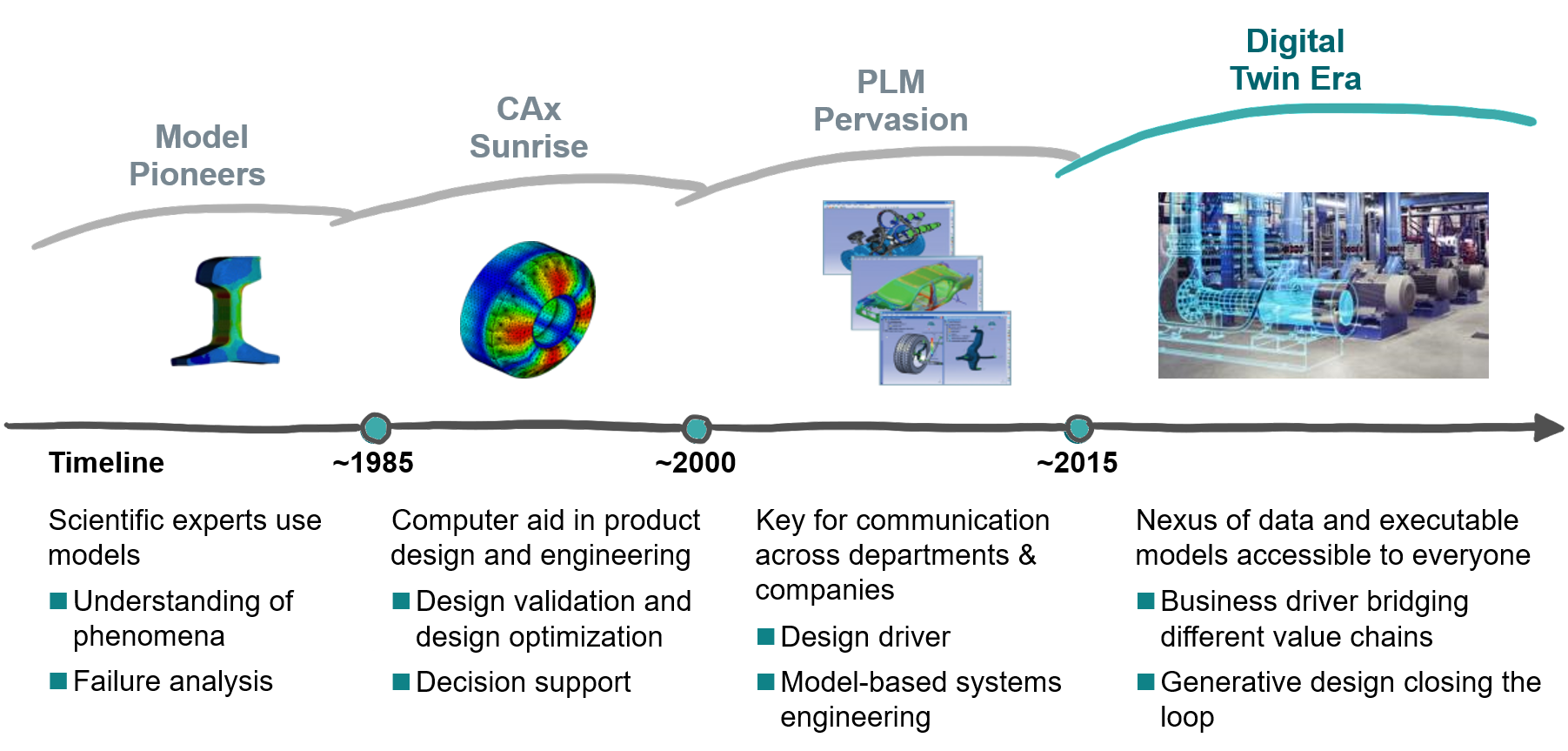}
    \caption{Waves of simulation}
    \label{fig:history}
\end{figure}

The concept of Digital Twins is not new. Actually the concept goes back to the early 60s (or even earlier) when simulation developed as a key tool for understanding scientific phenomena. For example, albeit many physical phenomena and basic equations have been understood, it was simulation which allowed to handle complex systems such as rockets and space ships. The interactions and geometries were far too complex to be solved by analytical means. Actually, still a number of today’s widely used simulation tools date back to these days, e.g. the NASA Structure Analysis software better known as NASTRAN\footnote{C.f. \url{https://github.com/nasa/NASTRAN-95} or the corresponding commercial versions}.

Simulation proved to be a very powerful if not the key tool to realize such complicated systems. Therefore, naturally simulation became a standard tool in many engineering disciplines\footnote{covering such divers fields as mechanics, electro-magnetic, fluid dynamics, thermodynamics, and many more} over the following decades. In particular, combined with optimizations simulations matured from a trouble shooting tool to standard virtual validation tools and ultimately design driving tools (Generative Design tools).

With mathematical models and their simulations developing as a success story for all engineering disciplines, it was a logical step to democratize them across the complete development process. With the rise of the so-called model based systems engineering, models became the backbone of systems engineering allowing always up-to date virtual validation on all levels from components to the system level validating product requirements. Purely document based systems were set down for model-based philosophies\footnote{It is actually an ongoing process and only very few examples exist with complex model-based engineering philosophies}.

\section{The Digital Twin Vision} \label{sec:digitaltwins}

With the evolution towards model-based (systems) engineering, the next logical step was to introduce the concept of Digital Twins extending model-based paradigms along the complete life-cycle. Early definitions of these concepts can be traced back to the works of \cite{grieves2014digital} or \cite{glaessgen2012digital}. But since then a multitude of different concepts and definition have been introduced, see e.g. \cite{rasheed2019digital}.

Digital Twins are expected to become a business imperative, covering the entire life-cycle of an asset or process and forming the foundation for connected products and services. Companies that fail to respond will be left behind. For example, it is predicted that companies who invest in Digital Twin technology will see a 30\% improvement in cycle times of critical processes \cite{pettey2017gartner}. A potential market of 90bn Dollar per year associated to corresponding offerings is predicted\footnote{\url{https://youtu.be/AtYEpvnEpp0}}, e.g. via savings or enabling novel business opportunities such as model-based diagnostics, prescriptive maintenance, or optimized operations. 
 
Let‘s us try to provide a very basic definition of Digital Twins:

\noindent\textbf{Definition (Digital Twin)}\footnote{For a short explainer see also \url{https://youtu.be/ObGhB9CCHP8}}\textbf{:}
The Digital Twin integrates all data (test, operation data,...), models (design drawings, engineering models, analyses, ...), and other information (requirements, orders, inspections, ...) of a physical asset generated along its life cycle that leverage business opportunities. The role of the Digital Twin is to predict and optimize performance. To this purpose simulation methods and / or data-based methods are used.\\

The Digital Twin itself however is only a central asset, if it can be used to make relevant predictions providing the right level of information at the right time. It bridges the physical and the virtual world and is a key tool to understand and model assets performance, predicting its behavior, and optimize it‘s operation and service. Since it spans the entire asset life-cycle continuous updating to mirror the reality is a central requirement. 

\subsection{Why Are Digital Twins Hot?}
As stressed above, the concept of Digital Twins is not new but the natural extension of model-based philosophies across the complete life cycle. However, with the exponential advancement of a number of key technologies, such as computing hardware, mathematical algorithms knowledge graphs and semantic technologies as well as augmented and virtual reality device, we are entering a new era. 

For example Moore‘s law has led over the past decades to an explosion of computational power \cite{schaller1997moore}. And actually a continuous capability growth  in computer and hardware architectures beyond scaling of chip performance and cloud is observed, e.g. graphics processing units (GPUs), re-configurable computing, ubiquitous computing \cite{arden2010more}. This is often referred to \textit{More-than-Moore}‘s law. At the sane time also major breakthrough in mathematics and computer science have been achieved. These have lead to an exponential capability growth contributing significantly to efficiency, scalability, and usability of (simulation) model-based as well as data-based approaches \cite{rude2016research}. However these enabling factors are often not recognized outside the Computational Science and Engineering community. Along with the growth of  computational and algorithmic capabilities also the interaction with digital models has been simplified. In particular, Virtual and mixed reality is entering professional market ultimately driving democratization of formerly expert centric tools \cite{bellini2016vrar}. Furthermore the advancement of semantic standards and knowledge graphs now entering rapidly the industrial domain allows for efficient realization of ontologies. Semantic information access has reached a new level beyond general purpose inter-operable standards \cite{woods2018knowledgegraphs}. This in particular reduces the required manual efforts for many tasks.
 
Thanks to these technology leaps, it is not anymore about faster and more accurate  simulations but rather we can rather democratize simulation across a wide range of applications domains and the complete life-cycle. Thus it is absolutely justified to have named Digital Twins in three consecutive years as one of the top 10 technologies \cite{cearley2017gartner,kerremans2018gartner,kerremans2019gartner}. Or to quote the ASSESS community: \textit{The Digital Twin approach will [..] make engineering simulation widely available appropriate to support improved decision making throughout the entire life-cycle of engineered products and processes}\footnote{https://www.assessinitiative.com/wp-content/uploads/Understanding-and-Enabling-the-Simulation-Revolution.pdf}.
 
\subsection{Digital Twins - Model- and Data-based Predictions}
 
From a high level point of view, information included in Digital Twins can be split in two categories, pure data values with only little additional structure / knowledge associated, such as data gathered from sensors, and structured executable model-based data, in particular simulation models. Thus, Digital Twins bring together classical data based schemes with model-based approaches such as simulation and optimization. Albeit the scientific communities today are still very separated, we believe that only a combination of the two will enable the success of Digital Twins. Albeit hybrid models combining the best of the two worlds (e.g. combining response-surface modeling with equation based models) are not new, an explosion in terms of publications can be observed in particular in the last years. 

\begin{figure}[h]
    \centering
    \includegraphics[width=15.0cm]{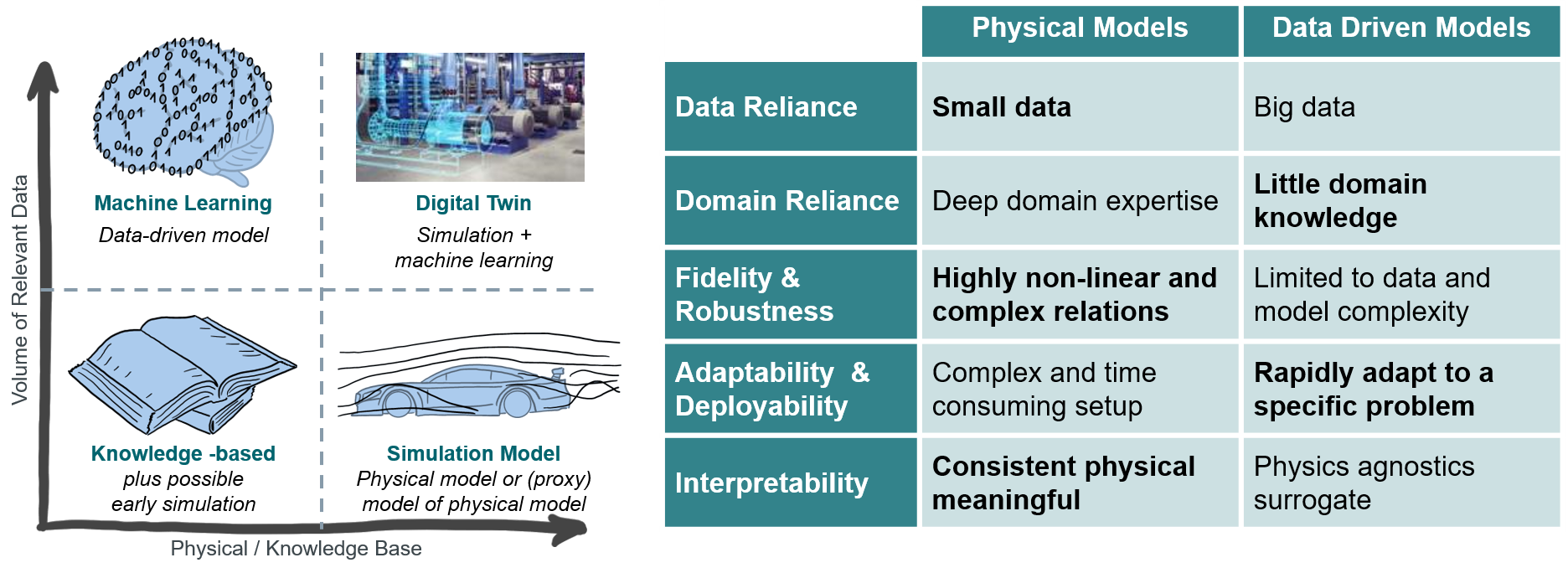}
    \caption{Digital Twins embrace Model-based as well as data-based approaches (Inspired by Lior Horesh)}
    \label{fig:hybridapproaches}
\end{figure}

\section{Digital Twins Enable the Impossible} \label{sec:examples}

As stated above, Digital Twins are key tools to address toady's industrial challenges. In many cases the offer a unique potential for higher efficiency and savings. For example in the case of predictive power trains the cost / efficiency ratio has been compared with different other measures \cite{haas2012predictive}. But very often it is only Digital Twins which enable a novel solution or service at all - often tasks which have been considered to be impossible. To be more concrete we will review four applications below.

\subsection{Interactive Design Tools}\label{subsec:interactivedesign}
Typical industrial design processes today require the coordination of experts with different specialization. For example, in the case of the geometric design of a component at least a designer, creating and adapting the geometric design within a CAD tool, and an analyst / simulation engineer, validating and optimizing the design by means of physical simulations in a CAE tool, are required. Their interaction is typically iteratively with each step requiring hours to sometimes days (in particular for the analysis step). The required efforts therefore limit the potential design configurations to be investigated.
Considering additive manufactured parts this is a severe limitation. Being constrained only by very few design constraints, the potential design space is quasi infinite. Thus the search for the optimum design is limited by the available time for the design process. Providing faster and more CAD-integrated analysis tools in particular for the very early concept design phase would allow to better exploit the potential design space. Not only would this reduce development costs but due to the opportunity to search even for more design options, this would at the same time also increase product performance.

Combining
\begin{itemize}
    \item \textbf{Efficient voxel-based discretization techniques} allowing lean codes, efficient memory access, easy parallelization, and fast transfer to novel compute architectures such as graphic processing units,
    \item \textbf{Geometric multi-grid solvers} enabling an optimal solution of linear systems in terms of complexity, and
   \item \textbf{Hardware-aware implementations} exploiting massively-parallel graphics processing units,
\end{itemize}
we have developed highly accelerated solvers ready for industrial use \cite{gavranovic2019topology}. Depending on the use case the solvers outperform state-of-the-art solutions by several orders of magnitude while having an appropriate good enough accuracy. In particular in the early design phase their are still a large number of uncertainties, such as the expected forces.

Potential applications range from accelerated topology optimization \cite{gavranovic2019topology} to sufficiently accurate 3d physics simulations in virtual reality\footnote{Design concepts in VR \url{https://youtu.be/ecg9JxsrNw8}} \cite{breuer2018idea}. Without sacrificing the achievable performance the concept phase can be accelerated up to 10 times. At the same time a much larger user base can be addressed with simulation tools leading to a democratization in terms of usage \cite{hanna2017democratization}.

\begin{figure}[h]
    \centering
    \includegraphics[width=12.0cm]{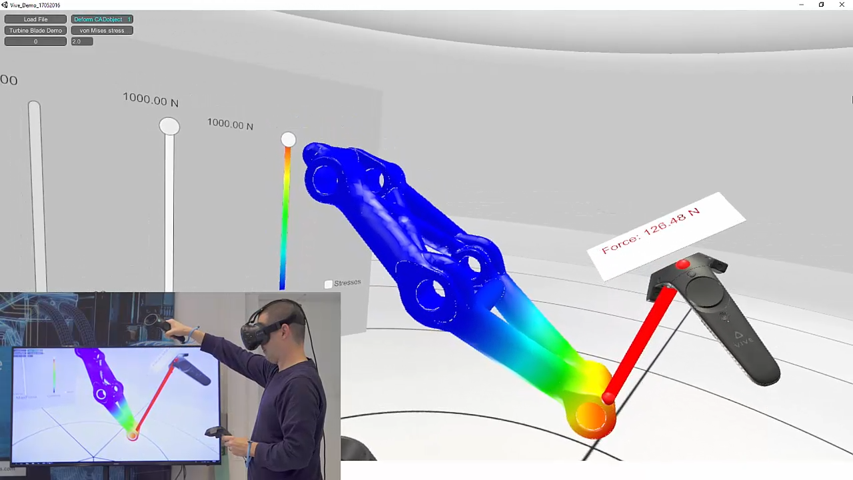}
    \caption{Interactive 3d multi-physics FEM simulation in a virtual environment}
    \label{fig:cae_in_VR}
\end{figure}


\subsection{Robot Milling}\label{subsec:robotmilling}
Albeit the advancement of additive manufacturing technologies many parts are still produced by subtractive manufacturing processes. Due to the high process forces in machining, in particular for metals, corresponding machine tools need to be extremely stiff such that they are not impacted by the high forces. Any deflection of the machine would lead to quality defects. Machining itself is typically split into two separate tasks: roughening removing the large part of the material and finishing of the surface. While roughening has accuracy requirements around 0.1mm, finishing often requires accuracies below 0.01mm. The corresponding high mechanical requirements are a key cost driver of machine tools.

Therefore, it is not surprising that machining has been considered as a use case for more flexible and versatile robots, e.g. robots could be also used for other tasks such as loading loading of machine tools and they have a significantly largest operating space while they are cheaper than machine tools. Today they are already used for soft material milling such as foams in industrial applications. The use for metal machining has not been considered realistic in terms of achievable accuracy.

Introducing
\begin{itemize}
    \item\textbf{Novel voxel-based milling process force models} based on fine-resolution multiscale geometry models combined with simple heuristic force models and
    \item\textbf{Model predictive compensation solutions} offered by the recent advancements in controls\footnote{Siemens SINUMERIK module CC ROCO-Robot Accuracy},
\end{itemize}
we could show that the error for aluminium parts could be reduced from around 1.0mm to 0.1mm using standard robots without any mechanical modifications \cite{schnoes2019robot}.

Digital Twins ultimately allowed to realize the application of robot milling\footnote{Precise, Digital and Flexible \url{https://youtu.be/2iIN-9Kno3o}} which was widely believed not to be possible satisfying industrial requirements for metal milling with conventional robots \cite{krull2018robots}.

\begin{figure}[h]
    \centering
    \includegraphics[width=12.0cm]{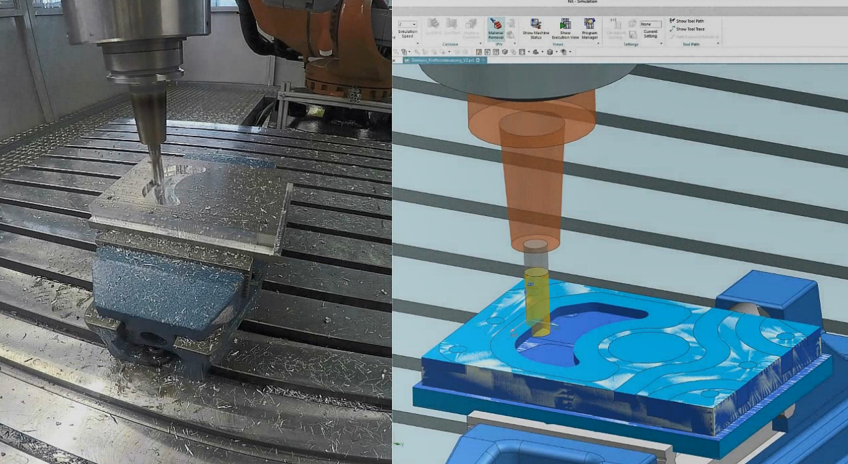}
    \caption{Milling robot in action together with the corresponding digital twin}
    \label{fig:robotmilling}
\end{figure}


\subsection{Robotic Cleaning of 3D Prints}\label{subsec:roboticcleaning}
Albeit additive manufacturing is a rather old technology it has only recently taken the barrier of industrialization. Among the many technologies, selective laser melting is one of the most widely adopted additive manufacturing technologies for metals. Thereby a powder bed is build up layer-wise and the different layers are welded together with a laser. The powder in between remains untouched and as such acts as a support structure for the next layer. Thus if the final part includes void volumes, these are filled with metal powered. 

Before the actual use of the parts the metal powder needs to be removed, in particular due to its potential noxious effects. Since the corresponding shapes can be highly complex (c.f. Subsection \ref{subsec:interactivedesign}) and are very often produced only in small lot numbers an automation of this process is not possible. Very often the cleaning of the additively manufactured parts is a manual process. 

Using a combination of 
\begin{itemize}
    \item \textbf{A highly simplified discrete element method} representing coarse grained models of the metal powder,
    \item\textbf{Heuristic optimization algorithms} based on the Fast Marching Method for an efficient determination of emptying strategies, and
    \item\textbf{Automatic control code generation} which can be transferred seamlessly to automated de-powdering machines,
\end{itemize}
we have realized a robotic depowdering machine which can highly autonomously clean 3d printed parts requiring only the basic geometric information of void volumes \cite{breuer2018following}. The reduction of manual work, in particular in combination with potential health issues, is a key building block for industrialization of additive manufacturing. In addition less powder is lost providing providing a significant amount of savings.

\begin{figure}[h]
    \centering
    \includegraphics[width=12.0cm]{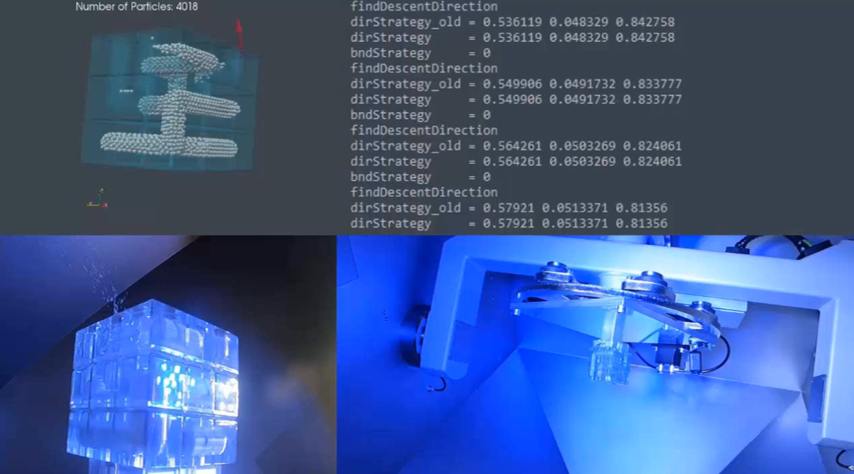}
    \caption{Operation of a depowdering machine and its corresponding digital twin}
    \label{fig:empty3Dprints}
\end{figure}


\subsection{Online 3D Virtual Temperature Sensors}\label{subsec:virtualsensors}
Many machines today are limited by thermal effects, i.e. could run to hot leading to sequent damage. For example, during startup large electrical machines can develop high temperatures at the rotor - limiting the operation. However, these can hardly be measured using cost-efficient sensor technology.  To prevent corresponding damage, they are often subject to conservative operation conditions avoiding overheating. Since stator temperatures can easily be measured, of course rotor temperatures can be estimated. However this requires significant manual efforts to build corresponding thermal networks and to validated them. Limiting their applicability for certain types of machines.

Combining
\begin{itemize}
    \item\textbf{Model order reduction} speeding up standard finite element models (FEM) from engineering by factors of up to 10,000, 
    \item\textbf{Continuous calibration in the loop} of these models combined with uncertainty combination to provide statistical information on the validity of the predictions, and
    \item\textbf{Immersed user experience} leveraging state-of-the-art mixed reality technologies with online simulations,
\end{itemize}
allow to 3d build virtual thermal sensors determining temperature distributions in the complete motor. By having efficient means to estimate critical rotor temperatures limiting operations today, higher availability of large electrical drives can be achieved. Having in mind that this can save costs of up to 200kEUR/h this is a tremendous achievement \cite{barnard2018virtual}. At the same time using model order reduction technologies \cite{hartmann2018model} the concept can be built directly on top of engineering models allowing to achieve the benefits with only little additional efforts. Mathematics enables quasi thermal x-rays for electrical motors allowing to monitor temperature distributions in real time\footnote{Virtual X-ray for large motors \url{https://youtu.be/86vkjykbHRM}}.

\begin{figure}[h]
    \centering
    \includegraphics[width=12.0cm]{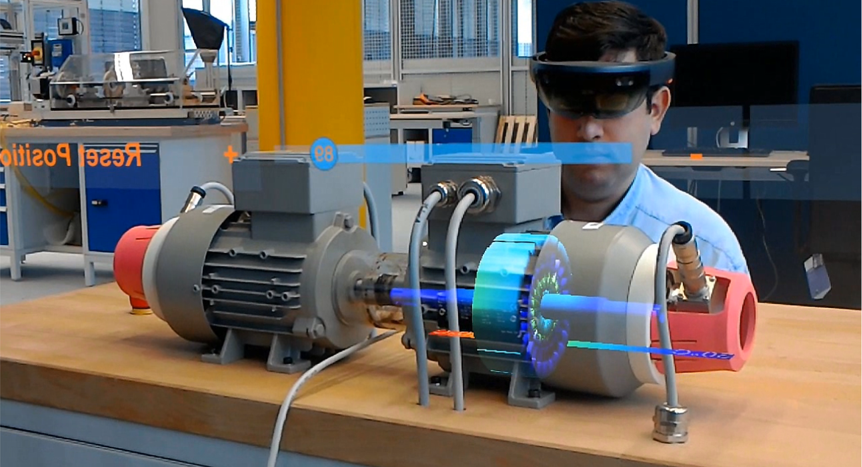}
    \caption{Mixed reality setup allowing to measure spatial temperature distributions parallel to operations by means of online simulations}
    \label{fig:motor_ar}
\end{figure}

\section{The Next Wave - Executable Digital Twins} \label{sec:executabledigitaltwins}
Albeit the previous section has shown four great examples of Digital Twins, a major limiting factor today is the manual work required to realize a Digital Twin, i.e. transfer of the corresponding models between different domains or life cycle phases. At the same time this requires to provide the models in the right execution environments with the right online capability in particular during the operations phase. 

We therefore introduce the concept of an Executable Digital Twin, which will be from our point of view a key aspect in any future Digital Twin driven application.\\

\noindent\textbf{Definition (Executable Digital Twin):}
An Executable Digital Twin is a specific encapsulated realization of a Digital Twin with its execution engines\footnote{Typically models today are distributed separately from their execution / simulation tools}. As such they enable the reuse of simulation models outside R\&D. In order to do so, the Executable Digital Twin needs to be prepared suitably for a specific application out of existing data and models. In particular it must have the right accuracy and speed. The Executable Digital Twin can be instantiated on the edge, on premise, or in the cloud and used autonomously by a non-expert or a machine through a limited set of specific APIs.\\

\begin{figure}[b]
    \centering
    \includegraphics[width=12.0cm]{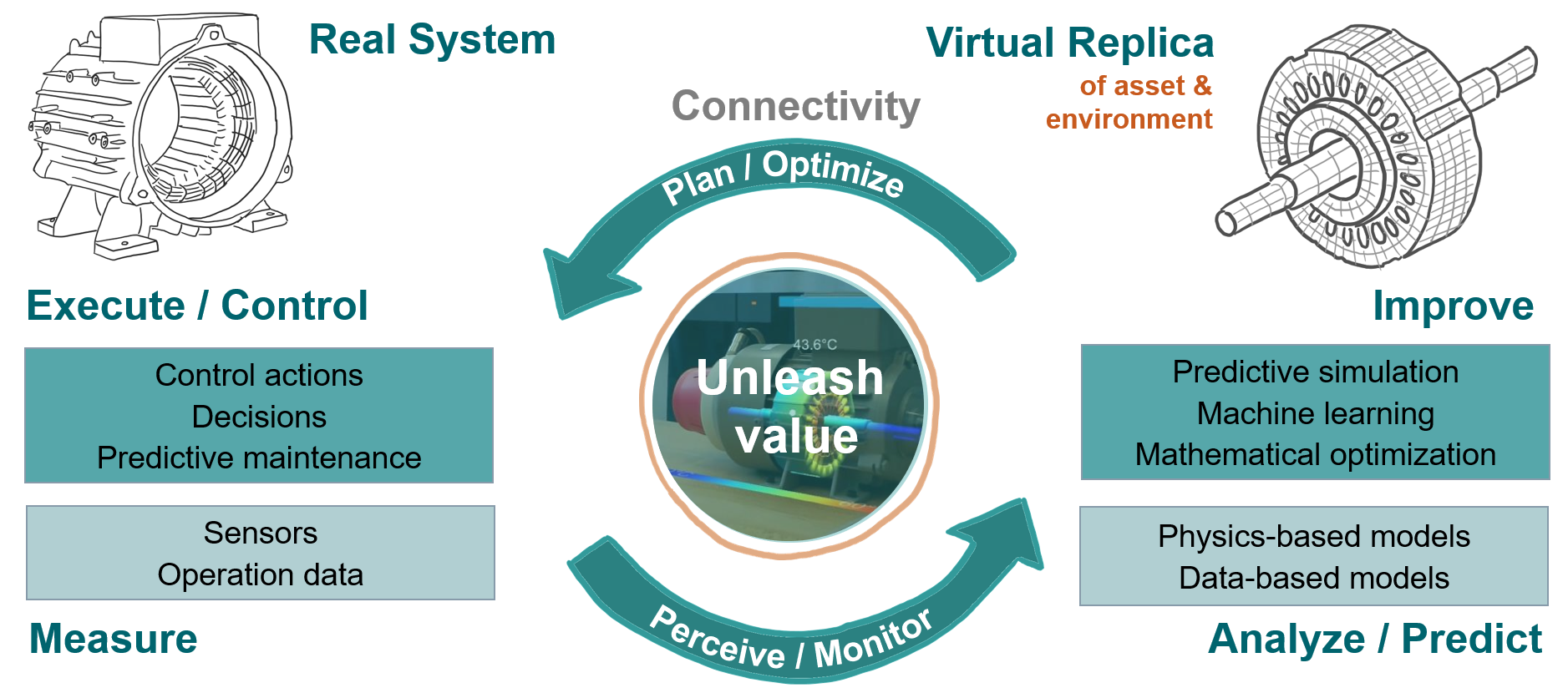}
    \caption{Operation parallel Executable Digital Twin (Inspired by Wim Desmet)}
    \label{fig:executableDT}
\end{figure}

Executable Digital Twins are a key enabler for democratization of Digital Twins. Their use can span a wide range of scenarios:
\begin{itemize}
    \item \textbf{System Integration:} Executable Digital Twins can be used for simulation models within a systems simulations. This could include scenarios of 3d solver coupling, the integration of 3d analysis within system simulations, or the integration of lumped system models. Thereby, the various models could originate from various parties in a complex ecosystem and also be based on different types of computational solvers. Applications can cover a wide range of scenarios, from sales support to sophisticated design and engineering tasks, such as virtual integration of a new component or subsystem, operating condition assessment, or performance limit evaluation.
    \item \textbf{Embedded Digital Twins:} Executable Digital Twins could be deployed on the edge within the embedded software stack to realize e.g. virtual sensors to enrich available information (c.f. Subsection \ref{subsec:virtualsensors}) or to be used directly in model predictive control loops, Furthermore, they also allow for simplified X-in-the-loop simulations or controller tuning during commissioning.
    \item \textbf{Companion Digital Twins:}  Executable Digital Twins could be further deployed on the edge, on premise, or in the cloud as an additional support in the context of complex Internet of Things (IoT) applications. This could span application fields from advanced monitoring and diagnostics, to asset optimization (e.g. via model predictive control), or predictive maintenance. Ultimately with the advancements of 5G, communication latency issues will be decrease allowing the convergence of the different compute locations from embedded to cloud.
\end{itemize}

In all these scenarios the creator and consumer of the Digital Twins will be separated, contrary to most applications of simulation today\footnote{In the context of Artificial Intelligence applications this separation is a fact already today.} Ultimately, the creator and consumer can be connected through complex Digital Twin ecosystems \cite{boschert2018next}. Thus Executable Digital Twins will ultimately lead to a democratization for simulation.

Today, most model-based approaches, and in particular simulation, are domain specific and mostly used during design and engineering. The core concept of the Digital Twin is to extend their usage along the complete life-cycle and to deliver new services providing the right information at the right place in an efficient way. Obviously this concept of the executable Digital Twin enforces sophisticated requirements in order to be successfully employed in standard industrial practice.\\

\noindent\textbf{Requirements:}
\begin{itemize}
    \item \textbf{Interactivity} - Speed and accuracy define the value of simulation and Digital Twins. Being very accurate, today's model and simulation approaches are extremely time-consuming. Speeding them up, while retaining the right level of accuracy, is crucial for extending the use of Digital Twins.
    \item \textbf{Reliability} - Users of Digital Twins cannot be expected to be sophisticated experts, like it can be expected during the use in design and engineering. Thus any prediction by the Digital Twin must be fail safe and / or provided along with confidence intervals such that no expertise is required to interpret the results or can be used autonomously, e.g. by controls. In particular Digital Twins must be self-aware and inform the user if they are used outside their validation and specification domain.
    \item \textbf{Usability} - Model-based and simulation tools are expert centric today. Their resources are limited and thus the use of corresponding tools today is limited by the availability. Therefore, any Digital Twin solutions must be accessible also for non-experts from a usability perspective. Similarly efforts for their generation should be reduced as much as possible.
    \item \textbf{Security} - Many business models based on the Digital Twin will require to exchange Digital Twins between different parties. Reverse engineering must be prevented, such that no intellectual property is lost.
    \item \textbf{Deployability} - Digital Twins will be used different from the place where they have been created, e.g. on customer premises, in the cloud, on controls. Thus deployment must be easy to reduce barriers and efforts.
    \item \textbf{Synchronicity} - Digital Twins link the real with the digital world. Since each asset is typically unique, Digital Twins must be uniquely adapted and keep synchronous with their real world counter parts. This includes adaptation of models according to wear as well as updating the models due to reconfiguration or service replacements.
\end{itemize}

Many of these requirements have not yet today appropriate technical solutions. We believe that they all require sophisticated mathematical concepts ranging from model order reduction (Interactivity) and uncertainty quantification (Reliability) to block chain concepts (Security). Without further mathematical research the vision of Digital Twins and in particular Executable Digital Twins will not be successful.\\

\noindent\textbf{Acknowledgements:}
Acknowledgements: The authors greatly acknowledge the contributions of Utz Wever, Stefan Gavranovic, and Hans-Joachim Bungarz for the use case of interactive design tools (Subsection \ref{subsec:interactivedesign}), Florian Sch\"oes, Michael Z\"ah, Sven Tauchmann, Birgit Obst, and Frank Fischer for the use case of robot milling (Subsection \ref{subsec:robotmilling}),  Christoph Kiener, Frank Fischer, and Meinhard Paffrath for the robotic cleaner (Subsection \ref{subsec:roboticcleaning}), and Utz Wever, Birgit Obst, Theo Papadopoulos, Wenrong Wen, and Philipp Stelzig for the virtual temperature sensors (Subsection \ref{subsec:virtualsensors}): In particular also see the corresponding publications. Furthermore, we greatfully acknowledge the support of the Innovation Fund of the Siemens AG enabling the transfer of unique mathematical building blocks to industrial innovations.

\nocite{*}
\bibliographystyle{plain}
\bibliography{references}

\end{document}